\documentclass{article}

\usepackage{arxiv}

\usepackage[utf8]{inputenc} 
\usepackage[T1]{fontenc}    
\usepackage{hyperref}       
\usepackage{url}            
\usepackage{booktabs}       
\usepackage{amsfonts}       
\usepackage{amsmath}       
\usepackage{nicefrac}       
\usepackage{microtype}      
\usepackage{cleveref}       
\usepackage{graphicx}
\usepackage[numbers]{natbib}
\usepackage{doi}
\usepackage{multirow}
\usepackage{subcaption}
\usepackage{tabularx}

\title{Reliability Analysis of Monitoring System for Extraterrestrial Habitat using CTMC and Empirical Evaluation\\}

\author{ \href{https://orcid.org/0000-0002-5387-1497}{\includegraphics[scale=0.06]{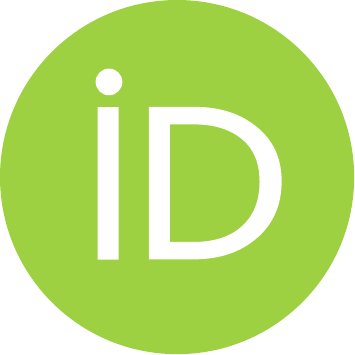}\hspace{1mm}Saurabh Band} \\
	Dept. Sustainable Communication Networks\\
	University of Bremen\\
	Germany\\
	\texttt{sband@uni-bremen.de} \\
	\And
	{Florian Stechmann} \\
	Center of Applied Space Technology and Microgravity\\
	University of Bremen\\
	Germany \\
	\texttt{stechflo@uni-bremen.de} \\
	\And
	{Malavika Unnikrishnan} \\
	Dept. Sustainable Communication Networks\\
	University of Bremen\\
	Germany \\
	\texttt{mal\underline{ }unn@uni-bremen.de} \\
	\And
	{Shadi Attarha} \\
	Dept. Sustainable Communication Networks\\
	University of Bremen\\
	Germany \\
	\texttt{sattarha@uni-bremen.de} \\
	\And
	{Christiane Heinicke} \\
	Center of Applied Space Technology and Microgravity\\
	University of Bremen\\
	Germany \\
	\texttt{christiane.heinicke@zarm.uni-bremen.de} \\
        \And
	{Andreas Willig} \\
	Dept. Computer Science and Software Engineering\\
	University of Canterbury\\
	New Zealand \\
	\texttt{andreas.willig@canterbury.ac.nz} \\
        \And
	{Anna Förster} \\
	Dept. Sustainable Communication Networks\\
	University of Bremen\\
	Germany \\
	\texttt{anna.foerster@uni-bremen.de} \\
}

\renewcommand{\shorttitle}{Reliability Analysis of Monitoring System for Extraterrestrial Habitat}

\hypersetup{
pdftitle={Reliability Analysis of Monitoring System for Extraterrestrial Habitat using CTMC and Empirical Evaluation},
pdfsubject={q-bio.NC, q-bio.QM},
pdfauthor={Saurabh Band, Florian Stechmann, Malavika Unnikrishnan, Shadi Attarha, Christiane Heinicke, Andreas Willig, Anna Förster},
pdfkeywords={Monitoring System, Space habitat, Life Support System, LoRa, Wireless Communication, Smart Home Automation, Failure Tolerance WSN, MAC for LoRa},
}

\begin{document}
\maketitle

\begin{abstract}
	Much research has been done to settle a civilization on Mars in recent years. Among the various required resources for this civilization, the habitat is one of the crucial resources to live on Mars. Such an extraterrestrial habitat is designed to provide a safe place to live during the initial missions. It is equipped with monitoring and life support systems to ensure the astronauts' safety. In this work, we present a robust monitoring system with a use case for extraterrestrial habitats. Similar to a typical monitoring system, it consists of sensor nodes and a gateway connected through a wireless communication channel. In our system, we introduce robustness to various failures that can occur after deployment, namely, board, sensor and gateway failure, in the form of redundancy. For each failure, the problem is tackled differently. For the first two types, we use additional hardware as backup, while for the last type, we use neighbouring devices as backups. The backup devices function as replacements for the failed component, which helps the system to collect the data which would otherwise be lost. We evaluate how much the performance of the system improves by using backup devices. We use a Continuous-Time Markov chain for the theoretical evaluation and an experimental setup that includes the hardware prototype for the empirical evaluation. We also analyze the effect of a simple medium access mechanism on the system's performance in the presence of heavy noise on the channel. Based on our requirements, we use a simple custom medium access control (MAC) algorithm called Slotted-ALOHA with Random Back-off (SARB) to make communication reliable. We demonstrate that around $30-34\%$ of the packets are recovered, with the use of backup devices (redundancy), which would otherwise be lost in case of failures. We also demonstrate that the system's performance improves by $3.8-13.2\%$ with the use of a simple medium access technique (SARB).
\end{abstract}


\section{Introduction}

Current monitoring systems focus on predictive maintenance, predicting failure in the system being monitored. Such systems are of great significance where failure might have fatal consequences. One such application is the habitats for extraterrestrial surfaces. However, the monitoring system itself is also prone to faults or failure. Even though such events are not frequent, debugging such faults is comparatively easier on Earth than on extraterrestrial surfaces. This is mainly because the debugging of wireless sensor networks is not yet fully automated, and human resources would be limited on Mars, especially during the initial missions. Even with ground-earth support, it would be difficult to deal with time-critical faults due to the delay in communication. In the case of Mars, the round-trip communication delay to Earth can be around $20\ minutes$. Thus, it becomes essential to have a monitoring system that is very robust to failures and has a self-diagnostic system that can detect faults without human interference.

The habitats for extraterrestrial surfaces need to withstand extreme environmental conditions. All the systems needed to support life are integrated into the habitat (E.g. Life Support System). Major faults in these sub-systems may potentially lead to the loss of crew. Thus only having a system to monitor these systems is not sufficient. It is also important that the monitoring system is robust to faults itself. In the extraterrestrial habitat, multiple components are subject to failure. Certain failures might affect the system partially, while some might put the system at a standstill. For example, if a sensor node fails, a certain area of the system has a blind spot, whereas if the server fails, the entire system will crash. However, in both cases, the end result is the loss of data and the inability of the system to perform monitoring. Thus it is important to understand various failures that might occur in the monitoring system deployed for extraterrestrial habitats and different measures that could be taken to ensure the availability of the system. 

In this work, we address various failures that can happen in a monitoring system deployed in a Mars habitat and demonstrate how to tackle these failures while keeping the system available. In the scope of this work, we do not focus on how to diagnose the faults but instead ensure that the system does not go into complete failure. We make the system robust by ensuring the availability of the system and recovering the lost data due to the failure of individual components or congestion on the channel. This is done by adding redundancy in different forms for different components in the system. For the sensor node failure, we deploy additional boards that act as backup without human intervention, while for gateway failure, we use neighbouring gateways as a backup. We demonstrate that less data is lost during failures with the help of the presented methods.  We evaluate the availability of the system by calculating the probability of complete failure of the system using the Continous-Time Markov Chain (CTMC). The  methods are validated experimentally with hardware prototypes. 

Since the monitoring system is designed for extreme applications like extraterrestrial habitats, it is better to evaluate the system for similar use cases. The Moon and Mars Base Analog (MaMBA) \cite{heinicke_mamba-concept_2020} is a baseline of an extraterrestrial habitat developed at the Center of Applied Space Technology and Microgravity (ZARM), University of Bremen. It is designed specifically for hardware testing, including various human-centric technologies like the Life Support System (LSS), the power system and the monitoring system for the habitat. So, we consider it as the use case for our work and adapt the system to the requirement of the MaMBA. The system described in the paper can be implemented for similar habitats or any other application with similar requirements.

The rest of the paper is organized as follows. Related works are discussed in Section \ref{sec_related_work}. A brief overview of MaMBA is given in Section \ref{sec_mamba}. The system architecture of a proposed method is explained in Section \ref{sec_architecture}. The mathematical modelling of the secondary (redundant) devices with the Markov Chain is discussed in Section \ref{sec_redundancy}. Implementation details and system validation are presented in Section \ref{sec_implementation} and \ref{sec_analysis} respectively, followed by the discussion in Section \ref{sec_discussion}, Future Scope and Applications in Section \ref{sec_futurescope}, and Conclusion in Section \ref{sec_conclusion}.

\section{Related Works} \label{sec_related_work}

\subsection{Failure Detection in WSN}
Monitoring systems based on Wireless Sensor Networks (WSN) are widely used for applications like environment surveillance \cite{mainwaring2002wireless} and smart homes \cite{ghayvat2015wsn}. A  WSN is an excellent option for monitoring systems due to its low power consumption and low-cost \cite{chiang2004architectures}. In addition, they come with some added benefits, the duplication of the nodes to ensure large-scale coverage, flexibility to incorporate new sensor nodes, and the relatively small size of the node \cite{szewczyk2004analysis}. 

Many studies focus on WSN monitoring systems with embedded machine learning strategies for improving the system performance \cite{vu2022practical, sharma2021machine}. The study \cite{liu2021low} proposes a "low-power failure detector" to detect the failure of a node due to insufficient power. \cite{szewczyk2004analysis} examines the performance of the WSN without the implementation of any acknowledgements or message re-transmissions to conclude that, given the lower congestion in the network, the rate of successful delivery of messages is satisfactory. 

Furthermore, the study \cite{rullo2019redundancy}  reviews various methods for fault tolerance in WSN to address failures in sensing and routing. The first method aims to develop a fault-tolerant system, which handles the breakdown of one or more components and does not imply an overall system failure. As described in \cite{rullo2019redundancy}, a common way of implementing a fault-tolerant WSN system is by replicating system logic and system functions using physical redundancy, leaving out the faulty node from the whole system. Re-transmission is the second technique for reliable communication and routing failure. After sending data, the sender node waits for an acknowledgement from the receiver node. Upon failure of the acknowledgement, the sender node assumes a packet loss and re-transmits the message \cite{mahmood2015reliability}. In some studies, \cite{park2004scalable, wan2002psfq}, a negative acknowledgement from the receiver conveys the message of missing packets to the sender. 

Since wireless sensors are also prone to failure, there is literature about different approaches and methods used to detect these failures in the system. \cite{5449649} talks about the reliable sensor network using a health vector to inform the node's status. The nodes can fail due to battery failure or bad communication, which neighbouring nodes will detect. It does not address how to tackle the situation if the node fails. Some methods are based on data for detecting faulty nodes in the network. \cite{8101845} proposes methods based on wavelet transform and cross-correlation to detect a fault in sensor nodes. \cite{5491263} uses Neural Network to detect the fault based on the data of the sensors. \cite{7848647} proposes a method to detect forest fires by deploying a network of sensor nodes that transfer all the data to a common sink node for processing. However, the author mentions that sensor nodes used to sense the data can face various failures, which makes it important to verify the status of the nodes before relying on the data to make the prediction. Thus, the authors also propose a method which allows the sensor nodes to check their self-status before transmitting the data. This method consists of comparing self-data with that of neighbouring nodes. Some other methods detect faulty nodes in widely distributed networks using the Markov chain model \cite{7918145} and Fuzzy logic \cite{8286384}.
However, all these methods focus only on detecting faulty nodes or failures in the network. In life-critical applications, just detecting the failure is not enough. We also need to act upon it to avoid data loss. Some methods tackle this problem with different approaches. One such method \cite{pignaton_de_freitas_handling_2011} proposes moving sensor nodes carried by UAV (Unmanned aerial vehicles) to replace the faulty sensor node. \cite{mohamed_fault_2008} uses redundancy to make the system fault tolerant but only in terms of communication. The sensor network is connected with wires, and wireless communication is used as a backup.

\subsection{Types of Faults}
Faults in typical sensor nodes can be broadly categorized into hard and soft faults. A hard fault represents the state where the sensor node is not functioning at all. This results in no communication with the sensor node. A soft fault represents the state where the node functions but with some error \cite{5491263}. The error can be in the data or the functioning of the node itself. Usually, any failure that occurs is very specific to each node and might not have any correlation with the fault in neighbouring nodes. However, certain failures can affect all the neighbouring nodes in the vicinity, and such failures are called common cause failure (CCF) \cite{macedo_dependability_2014}. The external events that cause failure in all the functioning nodes simultaneously lead to CCF. For example, an explosion in the facility damages all the nodes in the entire room.  

\subsection{Dependability Analysis} \label{sec_dependability_analysis}
Redundancy is one of the most effective solutions to deal with failure while keeping the system dependable \cite{sklaroff_redundancy_1976}. Introducing redundancy in the system allows us to replace the faulty device with a spare one without affecting the system's availability. This has been stated in literature as dependability analysis. The dependability of the system has been researched widely with the help of mathematical models like Fault Trees and Markov Chain models. Such models help estimate the system's dependency by measuring Reliability and Availability metrics. 

One such study is presented in \cite{macedo_dependability_2014}, which estimates the dependability requirement of the system by using the Markov chain to build a redundancy model. The authors estimate the improvement in the system's Mean Time To Failure (MTTF) with backup devices compared to a system without any backup devices. It is done by creating a redundancy model of the system. To design the model, authors assume that the system will be in a good state as long as at least one device is functioning, and only when all the devices fail will the system be in a bad state. The rate of transition from a good state to a bad state is called the failure rate ($\lambda$), and the rate at which the system can be repaired is called the repair rate ($u$). The author considers various failure scenarios like switching failure and standby failure and thus includes multiple failure rates accordingly. The Markov chain model in this work tries to estimate the impact of redundancy by evaluating the MTTF of the system. \cite{macedo_dependability_2014} concludes that MTTF improves by $33\%$ using a single spare device for each primary device. The failure rate ($\lambda$) is considered to be a constant rate ($1\times10^{-4}$), i.e. one every $10,000\ hours$.
\\\\

Meanwhile, redundancy also brings specific issues like the increased data processing cost, higher delay and bottleneck in the overall network, and significant energy depletion as discussed in \cite{ismael2021esrra}. It discusses the spatial redundancy reduction approach to account for some of these problems, without compromising the system's reliability. In our work, we focus on increasing reliability by introducing redundancy. Other aspects are not covered in the scope of this work. 

\section{Moon and Mars Base Analog Overview} \label{sec_mamba}
\begin{figure} [h!]
\centering
    \begin{subfigure}[b]{0.2\textwidth}
         \centering
         \includegraphics[width=0.8\textwidth]{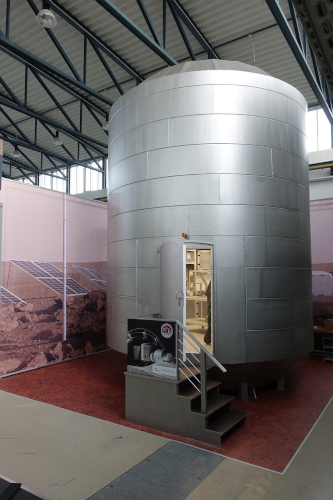}
     \end{subfigure}
     \begin{subfigure}[b]{0.2\textwidth}
         \centering
         \includegraphics[width=1.1\textwidth]{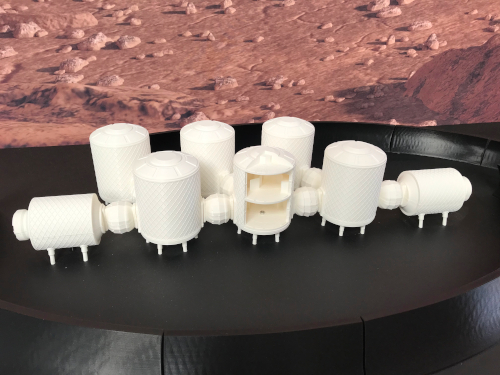}
     \end{subfigure}
\caption{The Moon and Mars Base Analog (MaMBA). (a) The exterior structure of the MaMBA habitat mock-up laboratory module \protect\cite{heinicke_mamba-concept_2020}, (b) The miniature model of the entire habitat with several other modules}
\label{fig_mamba_exterior}
\end{figure}

In this section, we give an overview of MaMBA. It consists of six upright cylindrical vessels designed to withstand low pressure. Out of the six modules, we focus on the laboratory module. For simplicity, we refer to it as 'the module' from here on. Figure \ref{fig_mamba_exterior} shows the exterior of the module's mock-up and the miniature version of the entire habitat. The module is a 2-storied structure where the lower part would be equipped as a laboratory and the upper part contains, by concept, a medbay and a workstation \cite{heinicke_mamba-concept_2020}. The inner diameter of the module is around $4.40\ meters$. This module is planned to be equipped with various components like a photobioreactor (PBR) that can generate oxygen for the crew, a power system to supply the module with power, and a communication system to deliver the sensor data from all the modules to the central processing server of the habitat \cite{heinicke_equipping_2021}. 

The safe running of all systems in a single module is ensured by monitoring various parameters. These parameters include temperature, humidity, air pressure, carbon dioxide, carbon monoxide, and oxygen. Along with the module, each subsystem can also be monitored. To monitor the PBR, various parameters must be observed in liquid and gaseous states. Dissolved oxygen, pH, temperature, and optical density are monitored for the liquid phase. Oxygen, carbon dioxide, temperature, relative humidity, and pressure are monitored for the gaseous phase. 

According to the system requirement of MaMBA, all the parameters need to be updated at least every \textit{40 seconds}; this interval is called \textbf{Maximum Monitoring Delay} (maximum time between two readings to ensure smooth and safe operation).

\section{System Architecture} \label{sec_architecture}

This section describes the architecture of the proposed monitoring system, which consists of three main components: sensor nodes, server (backend) and communication link between the sensor nodes and the server.

\begin{figure}[h!]
    \centering
    \includegraphics[width=0.45\textwidth]{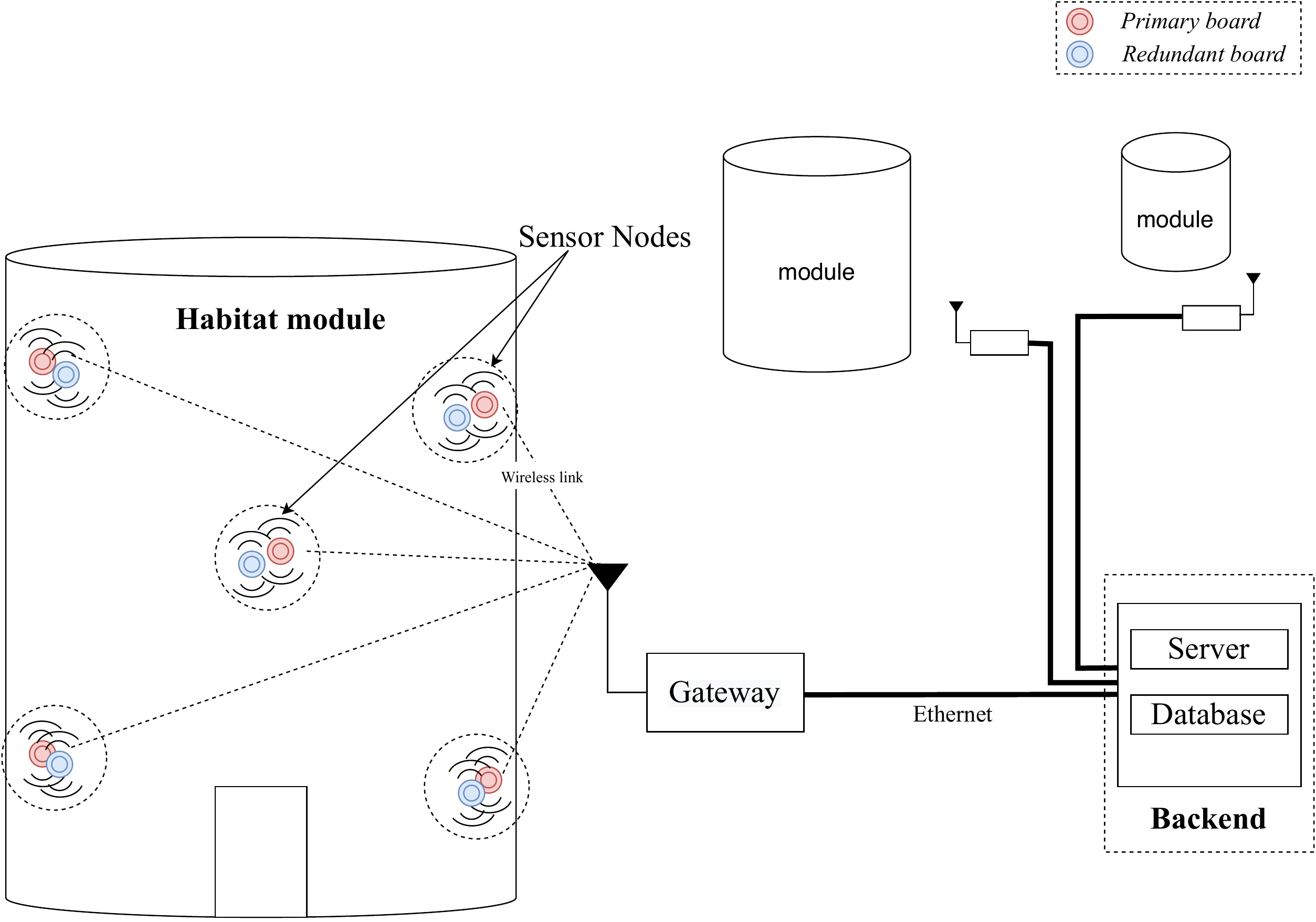}
    \caption{Generalized System Architecture of the monitoring system for extraterrestrial habitats.}
    \label{fig_system architecture}
\end{figure}

As seen in Figure \ref{fig_system architecture}, each module is monitored using multiple sensor nodes. The module represents the closed space to be monitored. Each module has a gateway to collect the data from the sensor nodes. The data from sensor nodes is transmitted periodically to the server via the gateway. The sensor node communicates with the server with converge-cast single-hop routing. The implementation of the MAC algorithm is discussed in Section \ref{subsec_communication}. The data processing takes place partially on the gateway and the server. Each sensor node comprises two physically identical sensing boards, each capable of sensing and transmitting the sensor data, called the primary board and secondary (backup) board. 

\begin{itemize}
  \item \textbf{Primary board (PB):} Main board of the sensor node, responsible for sensing. 
  \item \textbf{Secondary board (RB):} Replica of the primary board with identical monitoring scope. It is the backup to the primary board and only sends the data if the primary board fails
\end{itemize}

The two boards are denoted in Figure \ref{fig_system architecture} with blue (PB) and red (RB) colours. The rest of the paper will refer to the primary and secondary board pair as \textbf{sensor node}.  

The system can be adapted for any space habitat by choosing a concrete technology for each component. In this work, we demonstrate the system for the MaMBA space habitat use case. We will discuss the implementation details in Section \ref{sec_implementation}. 

\begin{table*}[htbp]
    \centering
  \caption{List of Sensors. }\label{tab_sensors}
    \begin{tabularx}{0.98\textwidth} {|
   >{\arraybackslash}m{3cm}|
   >{\arraybackslash}X |
   >{\arraybackslash}X|
   >{\arraybackslash}X|
   >{\arraybackslash}X|
   >{\arraybackslash}X|}
        \hline
        \textbf{Parameter}   &   \textbf{Component name}  &   \textbf{Unit}    &   \textbf{Measurement range}   &   \textbf{Interface}   &   \textbf{Quantity}    \\
        \hline
        CO$_2$      & SCD30             &   ppm     &     400 - 10000 ppm   &   I$^2$C      &   1           \\
        Ambient Pressure    &   BMP180  &   hPa     &     300 - 1100 hPa    &   I$^2$C      &   1           \\
        O$_2$       &   O2 SS Micro     &   percent &     0 - 25 \%         &   I$^2$C      &   1           \\
        CO          &   CO SS Micro     &  ppm      &     0 - 1000 ppm      &   I$^2$C      &   1           \\
        Temperature and Humidity &   AM2301          &   °C, \%  &     0 - 80 °C, 0 - 100 \%  &   one-wire protocol  &  4\\
        \hline
    \end{tabularx}
\end{table*}

\section{Redundancy} \label{sec_redundancy}

Based on the dependability analysis seen in Section \ref{sec_dependability_analysis}, we have seen that the MTTF of the system can be increased by the use of the backup devices, thus increasing the reliability of the monitoring system. However, it still does not address how likely this system is to fail completely. We evaluate the availability of the system by estimating the steady-state probability of being in a complete failure state.

As the system is made up of multiple components, the failure of any component can affect the reliability/availability of the system. Thus we use backup devices for each component, thus introducing redundancy in the system. This redundancy can be introduced by adding additional hardware which kicks in only when a failure occurs (sensor boards) or by using neighbouring devices that are already functioning (gateways). The approach to add redundancy can be different depending on the components. However, in all cases, the reliability/availability of the system should improve with each added backup device. 

\begin{figure}[h!]
    \centering
    \includegraphics[width=0.5\textwidth]{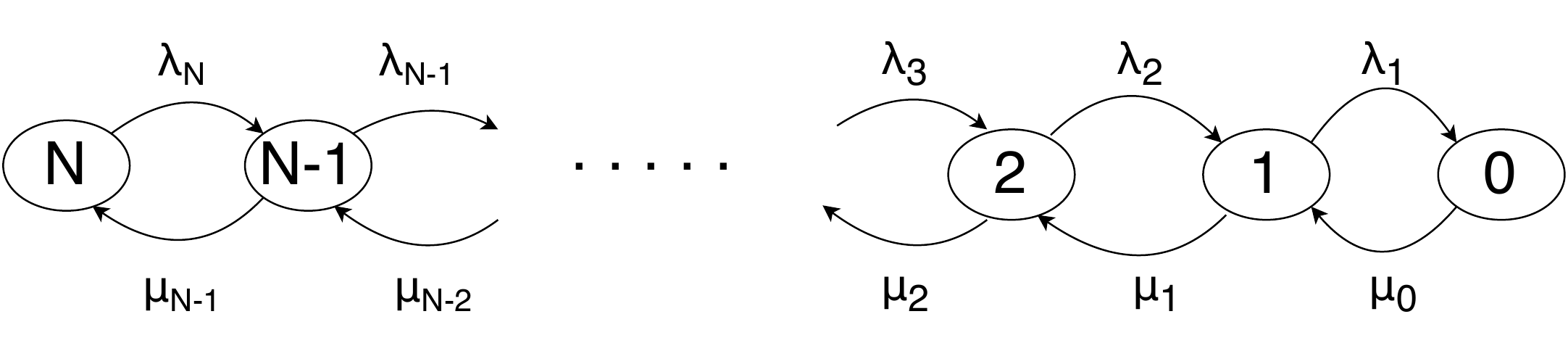}
    \caption{Generalized Markov Chain model for redundant system}
    \label{fig_gen_markovchain}
\end{figure}

Let us consider a single component of the system with multiple backup devices, i.e. a sensor node with multiple sensor boards. We model this using a Continuous-Time Markov chain with $N+1$ states having state space $S=\{0,1,2,......, N-1, N\}$, where $N$ represents the number of working boards. The state of chain $i$ at a given time is the number of working boards. For any state $i$, the failure rate is given by $\lambda_i$, and the repair rate is given by $\mu_i$ as shown in Figure \ref{fig_gen_markovchain}. If the chain is currently in state $i>0$, then a board failure will move it to state $i-1$. And if in state $i<N$ a repair occurred, then the chain will move in state $i+1$. Our resulting Markov chain is a birth-death chain as it can only transition to neighbouring states.  

We denote the generator matrix by Q as shown in equation \ref{eq_q_matrix}, where for any state $i$, the $i+1$ row of the generator matrix is given as $[\ \ ..., 0, 0, \lambda_i, -(\lambda_i+\mu_i), \mu_i, 0, 0,...\ \ ]$. If we define the steady-state vector for this CTMC by $\pi = (\pi_0, \pi_1, \pi_2, ..., \pi_N)$, then $\pi$ is the solution of the system $\pi\cdot Q = 0$ and the normalization condition is $\pi_0 + \pi_1 + ... + \pi_N=1$.

\begin{equation}
    Q = 
\begin{Bmatrix}
-\mu_0 & \mu_0 & 0 & 0 & .... & 0 & 0 & \\ 
\lambda_1 & -(\lambda_1+\mu_1) & \mu_1 & 0 & .... & 0 & 0 & \\ 
0 & \lambda_2 & -(\lambda_2+\mu_2) & \mu_2 & .... & 0 & 0 & \\ 
 & : &  & : &  &  &  & \\ 
 & : &  & : &  &  &  & \\ 
0 & 0 & .... & .... & 0 & \lambda_N & -\lambda_N & 
\end{Bmatrix}
\label{eq_q_matrix}
\end{equation}

For a birth-death chain, we can solve the above system for the steady-state probability of being in state $0$ by $\pi_0$ and other steady-state probabilities by $\pi_k$ ($k=1,...,N$) as shown in equation \ref{eq_p0} and \ref{eq_pk} respectively \cite{kleinrock_queueing_1976}. 

\begin{equation}
\label{eq_p0}
    \pi_0 = \frac{1}{1 + \sum_{k=1}^N \prod_{i=1}^k \frac{\mu_{i-1}}{\lambda_i}}
\end{equation}

\begin{equation}
\label{eq_pk}
    \pi_k =  \pi_0 \cdot \prod_{i=1}^k \frac{\mu_{i-1}}{\lambda_i}
\end{equation}

Assuming the general failure rate as $\lambda$, and also assuming that the boards are functioning independently, the failure rate for each state can be defined as $\lambda_i= i \cdot \lambda$. Assuming there is one repair person on-board who can repair one board at a time irrespective of how many boards are faulty, we can define the repair rate for each state as $\mu_i = \mu$, where $\mu$ is the general repair rate. By substituting these values in equation \ref{eq_p0}, we get the updated equation \ref{eq_p0_update} for the steady-state probability of the system in complete failure.

\begin{equation}
\label{eq_p0_update}
    \pi_0 = \frac{1}{1 + \sum_{k=1}^N \frac{\mu^{k}}{k! \cdot \lambda^{k}}}
\end{equation}

By substituting different values of $N$ in equation \ref{eq_p0_update}, we can calculate how likely the system is to fail with the different number of backup devices. We evaluate the system without any backup device, with backup devices, by substituting $N=1,2,3,4$.

Here, we only consider the general failure rate ($\lambda$), unlike \cite{macedo_dependability_2014}. Since there is no physical or electrical switching between the boards, we ignore the switching failure. In our case, as the secondary board constantly notifies us if it is working, we can also ignore the standby failure. However, we need to consider various failures that can occur as they would influence the repair rate. 

Broadly the failures can be divided into two categories; which could be repaired and which could not be repaired in the resource-constrained environment of the habitat. For failures that cannot be repaired, the best approach would be to calculate the mean time to failure as shown in \cite{macedo_dependability_2014}. On the other hand, the failures that could be repaired can be considered to occur at rate $\lambda = 1\times10^{-4}$, and repaired with rate $\mu=20.83\times10^{-3}$ (every $48\ hours$). 

As seen in Table \ref{tab_markov_results}, the comparative decline in $\pi_{0_{x}}$ with an increase in the number of backup devices, where $x$ represents the number of total deployed devices, demonstrates that the system becomes more reliable as we add secondary boards to the system. 
However, in the case of extraterrestrial missions, there are constraints on the weight and the mass of the payload to be carried. Thus we can say that using a single secondary node is a good trade-off between reliability and additional hardware cost.

\begin{table}[h!]
\centering
\caption{Failure probabilities of the system based on Markov Chain model}
\begin{tabular}{|l|l|l}
\hline
\textbf{No. of Nodes ($x$)} & \textbf{Failure Probability ($\pi_{0_x}$)} \\ \hline
1                     & $4.777e-03$                              \\ \hline
2                     & $4.565e-05$                              \\ \hline
3                     & $6.543e-07$                              \\ \hline
4                     & $1.250e-08$                              \\ \hline
\end{tabular}
\label{tab_markov_results}
\end{table}

The availability modelling with CTMCs, suffers from certain limitations, mainly that the times between events are always exponentially distributed. If we wanted to use other distributions (e.g. for the time until a component dies), we would have to approximate those, e.g. using the so-called phase-type distribution, which essentially is Markovian, but requires a much bigger state space.

\section{Implementation}\label{sec_implementation}

This section presents the implementation details of the proposed framework. The following subsections discuss the \textit{sensor nodes} and \textit{communication setup}.  

\subsection{Sensor Boards}

\begin{figure} [h]
\centering
\begin{subfigure}{.5\textwidth}
  \centering
  \includegraphics[width=0.5\linewidth]{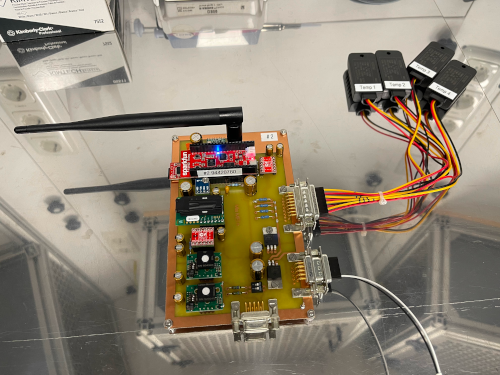}
  \caption{}
  \label{fig_board_hardware}
\end{subfigure}%
\caption{Prototype of the sensor board}
\label{fig_mamba_hardware}
\end{figure}

The individual sensor boards are equipped with eight sensors in the current prototype. As shown in Table \ref{tab_sensors}, temperature and humidity sensors have four instances, while all the other sensors have a single instance. The functioning of the board is handled by an ESP32 microcontroller. This microcontroller is booted with Micropython firmware. The prototype of the sensor board is illustrated in Figure \ref{fig_mamba_hardware}

The schematic diagram of the sensor board and working code can be found on the GitHub repository\footnote[1]{\href{https://github.com/saurabh-2905/sensorboard-mamba}{github.com/saurabh-2905/sensorboard-mamba}}. SPI and I2C protocols are used for onboard communication. Since the working voltage for ESP32 ranges between 2.2V to 3.6V, there is a voltage level converter on board to interface the sensors that work with voltages different than that of the controller. Based on the functionality, the boards are categorized as \textit{primary board} and \textit{secondary board}.

\subsubsection{Primary Board}
Out of the two boards in the sensor node, the primary board is responsible for sensing. The board senses the data from all the sensors continuously. It allows the system to detect any changes in the environment quickly. However, the data is transmitted at a lower rate than sensing in the form of packets. The transmission of the packets is handled by the MAC layer, which is discussed in Section \ref{subsec_communication}. For each sensor, pre-defined upper and lower thresholds define the normal values. If the detected values cross any threshold, it indicates an emergency and the packets are transmitted without delay. The rate at which packets are transmitted is called \textbf{transmission interval}. In a regular scenario, the packets are transmitted according to transmission interval. 

\subsubsection{Secondary Board} \label{subsec_redun_board}
\begin{figure}[h!]
    \centering
    \includegraphics[width=0.48\textwidth]{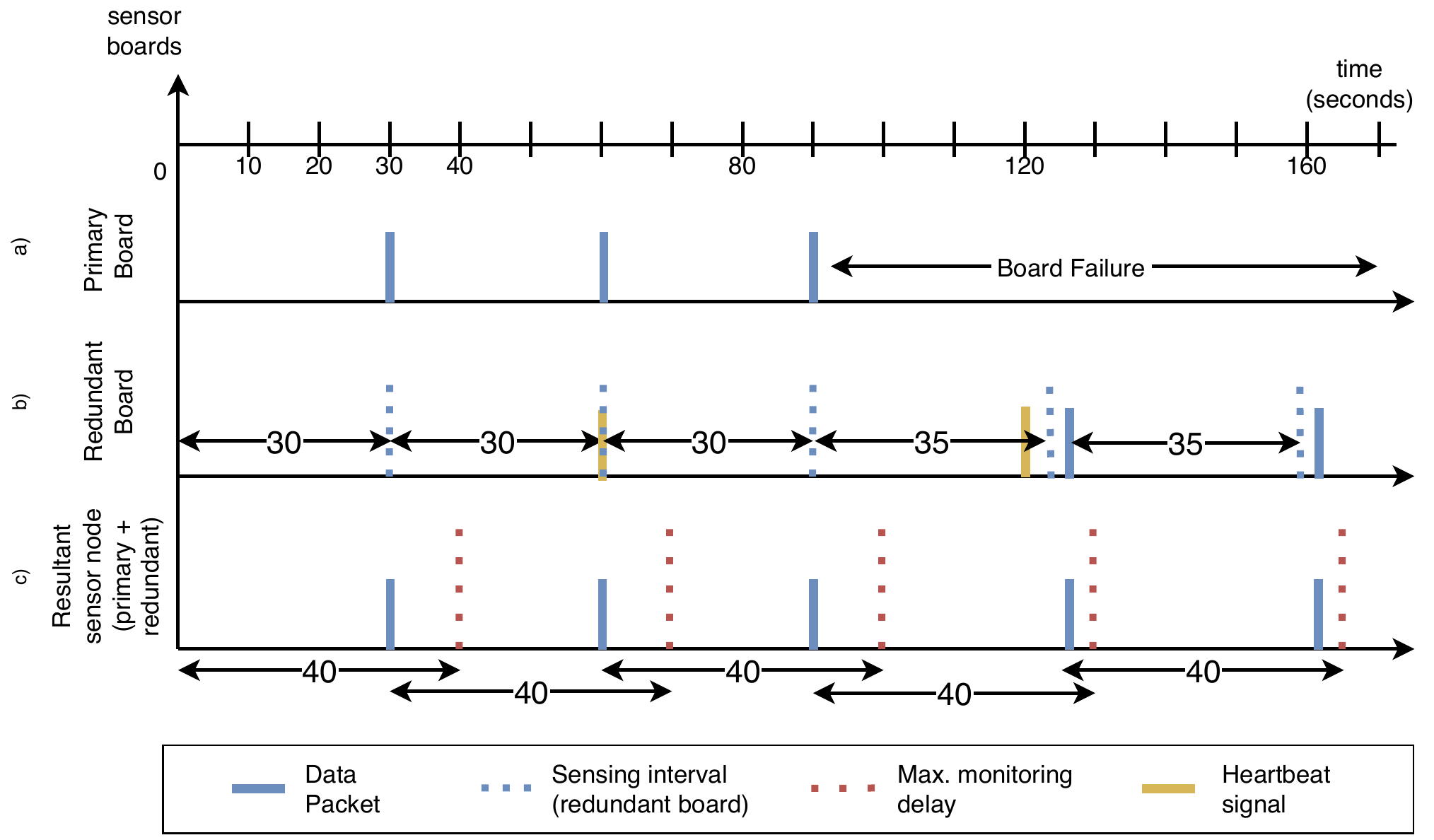}
    \caption{Packet reception at the server to demonstrate the redundancy concept. a) shows the packet transmission of the primary board. For simplicity, the transmission is considered constant at 30 seconds (max. transmission interval). b) shows the working of the secondary board. c) packet reception at the server from the sensor node.}
    \label{fig_redundant_diag}
\end{figure}

The secondary board is a backup for the primary board. A secondary board sends the sensor data packets if the primary board fails to do so; otherwise, while it is idle, it sends a smaller packet to indicate it is alive, known as \textit{heartbeat signal}. This function is explained in Figure \ref{fig_redundant_diag} with a timing diagram. The diagram contains three plots: $a),\ b)$, and $c)$, which show packet transmission of the primary board, secondary board and sensor node (primary + secondary), respectively. For better understanding, the transmission interval of the primary board in plot a) is assumed to be constant at $30\ seconds$. The secondary board waits for a few seconds more than the maximum transmission interval to check if the primary board is working; this interval is called \textbf{sensing interval}. 

Plot \ref{fig_redundant_diag}.$c)$ shows the resultant packets received from the sensor node and the maximum allowed delay for each packet (red dotted line). Plot \ref{fig_redundant_diag}.$a)$ shows the packets received from the primary board with $30\ seconds$ transmission interval. The secondary board will not send the packet as the primary board transmits the first three packets before the \textit{sensing interval}. However, it will send the heartbeat signal (yellow bar) every $60\ seconds$. After the third packet, as the secondary board does not sense any packet from the primary board till completion of \textit{sensing interval}, it will send a packet immediately. The difference between the completion of the sensing interval and packet transmission denotes the time required to read the sensor data. Even after the failure of the primary board, the packet from the secondary board reaches the server within the \textit{maximum monitoring delay}. The same phenomenon is seen for the next packet as well. As soon as the primary board is online, the secondary board stops transmitting packets and returns to transmitting $heartbeat\ signal$. One thing to note is that the secondary board does not have a re-transmission mechanism. It will transmit only once after it detects any packet is not transmitted by the primary board. Thus if the packet from the secondary board is lost, it cannot be recovered. As the secondary board is designed to compensate for the lost/corrupted data from the primary board in order to meet the time constraint, it would not be optimal to have a re-transmission mechanism for the lost packets. The re-transmitted packets from secondary boards would be delayed and would also result in an increase in traffic.

\subsection{Communication} \label{subsec_communication}
The wireless technology for communication was selected by comparing various technologies shown in Table \ref{tab_wireless_comparison}. Other popular technologies like Bluetooth, WiFi and Zigbee operate in the most crowded frequency band (2.4 GHz) and thus are more susceptible to interference. While other technologies like Zigbee, Z-wave and EnOcean work at lower frequency bands, the range is short compared to LoRa. Moreover, the compatibility of LoRa wireless technology in outer space is established in several other studies \cite{wu2019research}. The configuration parameters of the LoRa node can be modified to obtain optimized reliability as discussed in studies. Table \ref{tab_lora_p} shows the selected values for respective LoRa parameters based on literature \cite{yim_experimental_2018, cattani2017lora, ayele2017performance}. 

As LoRa PHY provides the radio interface without a medium access control (MAC) layer, the packets from multiple boards can collide at the gateway receiver. In \cite{pham_investigating_2018, rochester_lightweight_2020}, authors propose MAC algorithms with sensing capability like lightweight carrier sense (LCS) and CSMA/CA for LoRa. However, these methods include additional hardware and modified libraries to perform sensing. Even with the built-in sensing mechanism on the LoRa chips, Channel Activity Detection (CAD), it is not possible to access it via standard drivers for transmission. The other advanced MAC algorithms either try to save energy by waking the nodes only when needed or improve throughput for the busty traffic by combining TDMA and FDMA for multichannel communication \cite{huang_evolution_2013}. In short, all the random access techniques need to listen to the channel to avoid collisions. In contrast, the time-based access techniques need very high synchronization between nodes, a different research topic in itself. In addition, we do not want our nodes to sleep to save energy because this would mean the system is sensing at a smaller rate, making it less reactive to changes in the environment. We need a mechanism that would allow us to deliver the packet successfully with less collision and overhead. Thus we use a simpler MAC implementation for LoRa without channel sensing called Slotted ALOHA with Random Backoff (SARB). However, this algorithm does not need precise time synchronization as the slots are not pre-defined, which is one of the advantages.

\begin{table*}[t!]
\caption{Comparison of different existing wireless technologies. }
\begin{tabularx}{0.98\textwidth} {|
   >{\arraybackslash}X|
   >{\arraybackslash}X |
   >{\arraybackslash}X|
   >{\arraybackslash}X|
   >{\arraybackslash}X|
   >{\arraybackslash}X|
   >{\arraybackslash}X |}
\hline
& \textbf{Bluetooth}   & \textbf{Zigbee} & \textbf{Z-wave}   & \textbf{ENOCEAN}                                                           & \textbf{LoRa}   & \textbf{Wi-Fi}        \\ \hline
\textbf{\begin{tabular}[c]{@{}l@{}}ISM \\ Frequency\\ bands\end{tabular}} & 2.4GHz                                                     & \begin{tabular}[c]{@{}l@{}}2.4 GHz, \\ 915 GHz, \\ 868 MHz\end{tabular}  & \begin{tabular}[c]{@{}l@{}}2.4 GHz, \\ 908.4 MHz, \\ 868.4 MHz\end{tabular} & \begin{tabular}[c]{@{}l@{}}315 MHz, \\ 868 MHz, \\ 902.87 MHz\end{tabular} & \begin{tabular}[c]{@{}l@{}}433 MHz (Asia),\\ 868 MHz (Europe),\\ 915 MHz \\ (North America)\end{tabular} & \begin{tabular}[c]{@{}l@{}}2.4 GHz, \\ 5 GHz\end{tabular} \\ 
\textbf{Range} & 1 - 10 meters  & 1 - 100 meters  & 30 meters  & 30 meters   & 10 Km  & 1 -100 meters  \\ 
\textbf{Data rate}  & \begin{tabular}[c]{@{}l@{}}1 Mbps (v1.2)\\ 3 Mbps (v2.0)\\ 24Mbps (v3.0)\end{tabular} & \begin{tabular}[c]{@{}l@{}}250 kbps, \\ 40 kbps, \\ 20 kbps\end{tabular} & \begin{tabular}[c]{@{}l@{}}120 kbps \\ (868.3 Hz)\end{tabular}              & \begin{tabular}[c]{@{}l@{}}120 kbps \\ (868.3 Hz)\end{tabular}             & \begin{tabular}[c]{@{}l@{}}250 - 5470 bps\\ (depends on SF)\end{tabular}                                 & \begin{tabular}[c]{@{}l@{}}11 - 65 - \\ 450 Mbps\\ (IEEE 802.11n)\end{tabular} \\ 
\textbf{\begin{tabular}[c]{@{}l@{}}Nodes \\ per cluster\end{tabular}}     & 8  & 255& 255 &255 & 1000  & 32 \\ 
\textbf{\begin{tabular}[c]{@{}l@{}}Bandwidth \\ per channel\end{tabular}} & 1 MHz  & 2 MHz  & 100 KHz   & 280 KHz  & 125 KHz & 22 MHz \\ 
\textbf{\begin{tabular}[c]{@{}l@{}}Transmit \\ power\end{tabular}}        & 4 dBm& 0 dBm& 3 dBm & 6 dBm  & 13 dBm& 20 dBm  \\ 
\textbf{\begin{tabular}[c]{@{}l@{}}Network \\ Topology\end{tabular}}      & \begin{tabular}[c]{@{}l@{}}Star, \\ Peer-to-Peer\end{tabular}   & \begin{tabular}[c]{@{}l@{}}Star, \\ Peer-to-Peer, \\ Mesh\end{tabular}   & Mesh                         & \begin{tabular}[c]{@{}l@{}}Star, \\ Peer-to-Peer, \\ Mesh\end{tabular}    & Star& \begin{tabular}[c]{@{}l@{}}Star, \\ Peer-to-Peer\end{tabular}  \\ \hline
\end{tabularx}
\label{tab_wireless_comparison}
\end{table*}

\begin{table}[h!]
    \centering
    \caption{Selected values for LoRa parameters}
    \begin{tabular}{ |>{\arraybackslash}m{3cm}| >{\arraybackslash}m{3cm} |}
        \hline
        \textbf{Parameter}           &   \textbf{Selected value}  \\
        \hline
        Frequency           &   868 MHz       \\
        Bandwidth           &   125 kHz         \\
        SF                  &   7               \\
        CR                  &   4/5               \\
        \hline
    \end{tabular}
    \label{tab_lora_p}
\end{table}

For SARB, similar to slotted ALOHA, the packets are sent at specific time slots. Here, the slots are defined according to the transmission interval duration. However, we introduce a random back-off mechanism to make these time slots variable. As we want each packet to be delivered at the server within $40\ seconds$ from the previous packet, we define the range for transmission interval as $20-30\ seconds$. The transmission interval can vary with a step size of $500\ ms$. This interval varies randomly for each sensor node.

\begin{figure}[h!]
    \centering
    \includegraphics[width=0.49\textwidth]{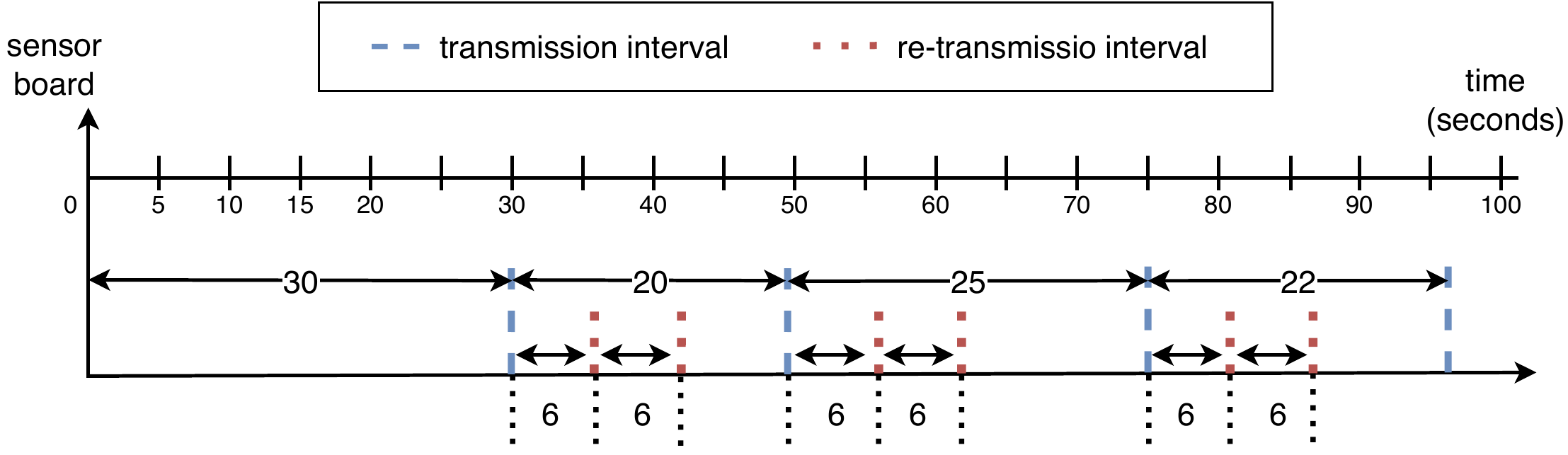}
    \caption{Visualization of variable transmission interval and re-transmission slots. }
    \label{fig_saba_queue}
\end{figure}

Figure \ref{fig_saba_queue} shows the transmission mechanism. There are two re-transmission slots between two consecutive transmission slots. We define the re-transmission interval as $6\ seconds$. Re-transmission is handled by a message queue that stores up to 10 packets. The packets in the queue are accessed according to the Last-In-First-Out (LIFO) rule. If the queue is full, the oldest packet is deleted to accommodate the new packet. All the intervals in the system can be defined by the developer based on the system requirements.

\section{Performance Analysis} \label{sec_analysis}

In this section, we validate the robustness of the system to various failures. To validate the system, we use the real hardware prototype of the sensor system. Each experiment contains a scenario that represents a failure of a certain component in the system and measures the reliability of the system due to the introduced redundancy. The scenarios tested are a) Hard Failure (HF) - complete failure of the primary board, b) Sensor Failure (SF) - failure of individual sensors leading to incomplete data, c) Gateway failure (GWF) - failure of the gateway leading to bottleneck. Since the MAC layer is selected to suit the given application, we also test how it affects the system's performance. All the experiments were performed for three iterations; the presented results are mean values of 3 iterations.

\textbf{Setup:} The setup used for this scenario contains one primary board, one secondary board, one noise board, one gateway, and one server. Failures are assumed to be outside the group of common cause failures, which affects only the primary board. We transmit a random signal on the channel every $500\ ms$ to simulate the channel noise. To put things in perspective, for our payload $(76\ bytes)$, each node has an on-air time of $138.5\ ms$. The noise signal of $10\ bytes$, transmitted every $500\ ms$, sums up to the on-air time of $494.4\ ms/minute$. This is equivalent to the transmission of $3\ boards/minute$ approximately. So if we consider a constant transmission interval of $30\ seconds$ for simplicity, the noise transmitted on the channel is approximately equivalent to $90$ active sensor boards. The experiments' duration is $30\ minutes$, out of which respective faults in each scenario are simulated for $20\ minutes$ out of the total time. The metrics used for evaluation are PRR (Packet Reception Ratio) and RSSI (Received Signal Strength Indication). 

\textbf{a) Hard Failure (HF):} This scenario simulates the complete failure of the board. This is done by switching off the power to the board, which is similar to battery failure in the real world, one of the common reasons for the failure of nodes. Normally, in this case, no data would be received until the battery is replaced and the board is back online. This creates a blind spot in the monitoring scope. 

\textbf{b) Sensor Failure (SF):}
This scenario assumes that the primary board is functioning and is able to send the data, but one of the sensors has failed. In the case of sensors, we can have two types of failures. First, the sensor cannot read and generate data, and second, the data generated by the sensor is incorrect (aka anomalous). The secondary (backup) board can detect both failures. Here we assume the backup board is working perfectly fine and has no faults. To verify the behaviour of the system in case of these sensor failures, we perform two different experiments. 

In the first experiment (SF1), we simulate the sensor's failure where the sensor cannot read. This can happen due to ageing or some external factors that damage the physical connection of the sensor. In the second experiment (SF2), we simulate the anomalies in the sensor data. If the sensor is damaged or not calibrated properly, it will generate wrong data (anomalous). We tackle this problem by comparing the data from the primary and secondary boards. If the data from both boards have a discrepancy greater than $25\%$, then we say the data is faulty. This mechanism can only detect if the problem exists but cannot localize which board is generating the wrong data. To answer this, more sophisticated methods should be used to detect anomalous data with the help of machine learning models \cite{attarha2023automated}.

\textbf{c) Gateway Failure:}
Data from all the sensor nodes pass through the gateway. Thus, it becomes the system's bottleneck, and the entire system will collapse in case of failure. Instead of adding a backup device similar to sensor nodes, we add redundancy in another form. As we use LoRa, which has a long range and can penetrate metal walls due to spread spectrum modulation, we can use the neighbouring gateways as backup receivers. The server can de-duplicate the packets that are received from multiple gateways. This aspect could not be considered in the case of other technologies without spread spectrum modulation.  

We verify this aspect with a similar experiment as before. Here we assume the sensor nodes to be working without any fault. Unlike the previous experiments, here, the gateway is placed outside the habitat at a distance of $12\ meters$. This scenario represents the gateway of the neighbouring module in real life scenario.

\begin{figure}[h!]
    \centering
    \includegraphics[width=0.5\textwidth]{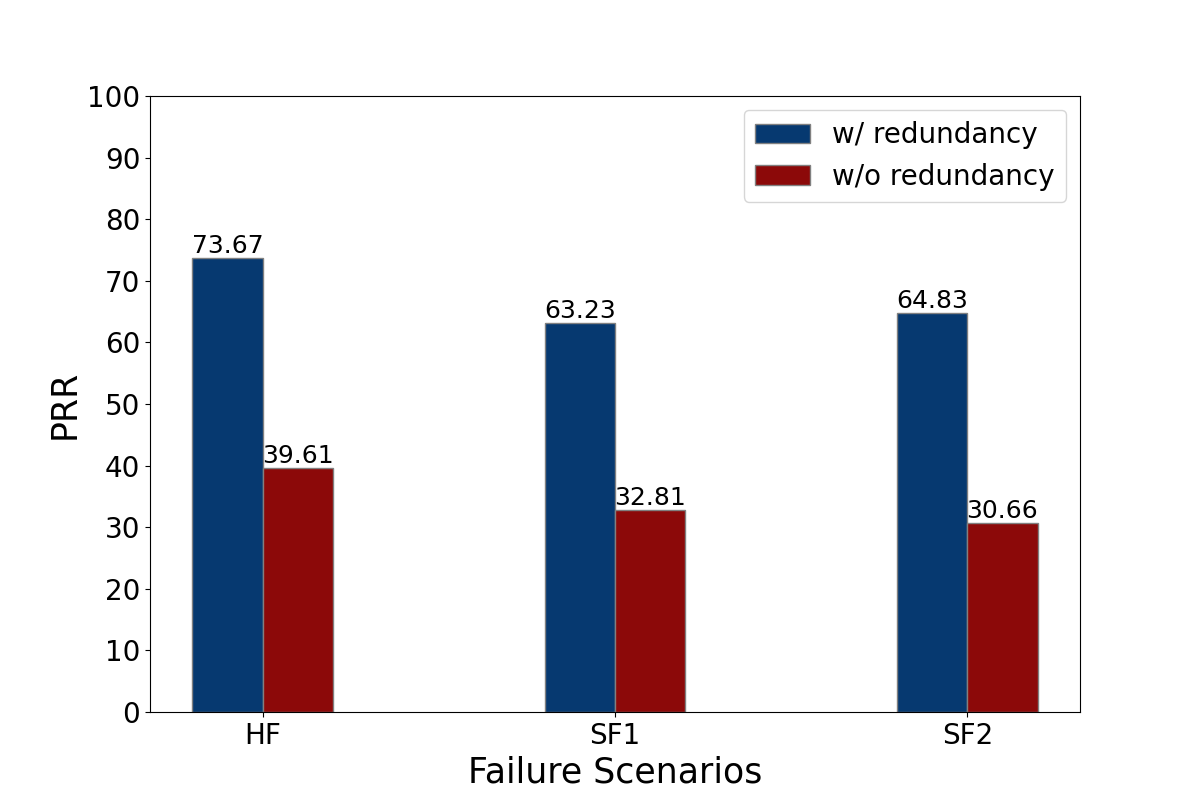}
    \caption{The PRR by sensor nodes during various failures, with and without redundancy. HF, SF1, and SF2 stand for Hard Failure, Sensor Failure Experiment 1 and Sensor Failure Experiment 2 (sensor anomalies), respectively}
    \label{fig_redundancy_prr}
\end{figure}

\begin{figure}[h!]
    \centering
    \includegraphics[width=0.5\textwidth]{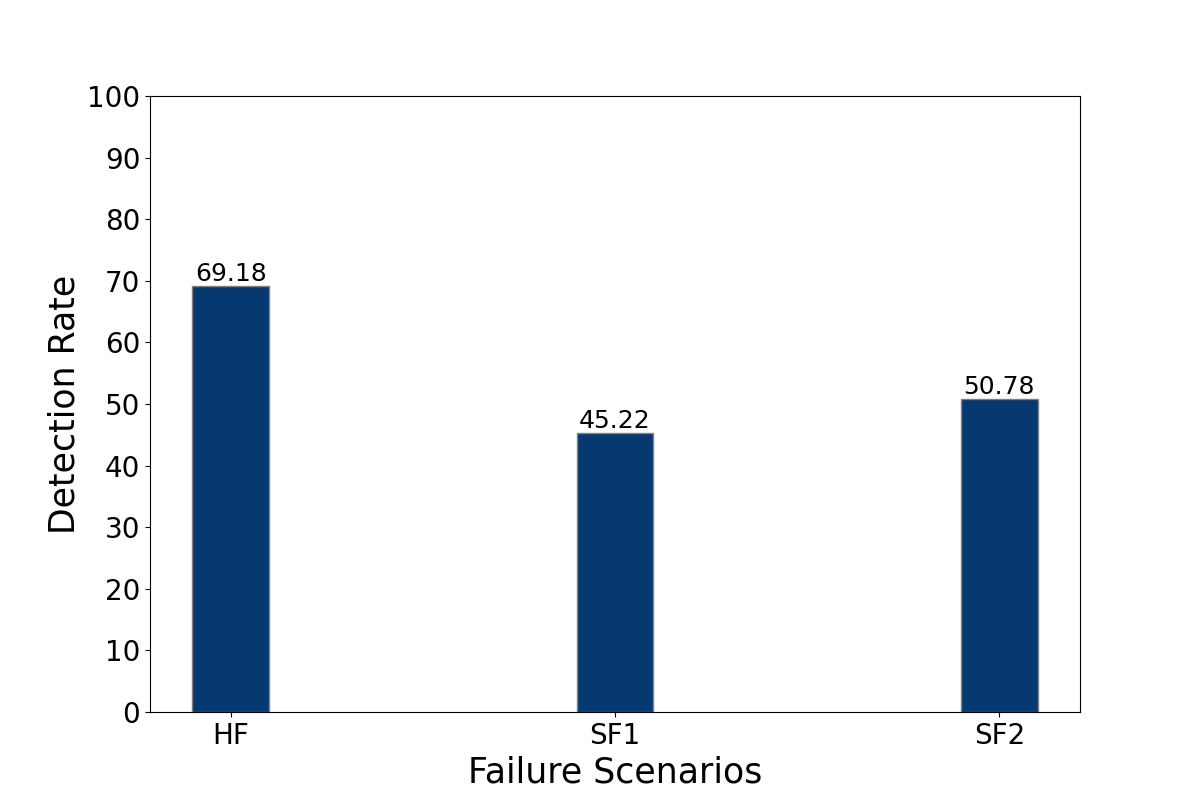}
    \caption{Detection rate of the secondary board for different experiments. HF, SF1, and SF2 stand for Hard Failure, Sensor Failure Experiment 1 and Sensor Failure Experiment 2 (sensor anomalies), respectively}
    \label{fig_detection_rate}
\end{figure}

\begin{figure}
     \centering
     \begin{subfigure}[b]{0.55\textwidth}
         \centering
         \includegraphics[width=\textwidth]{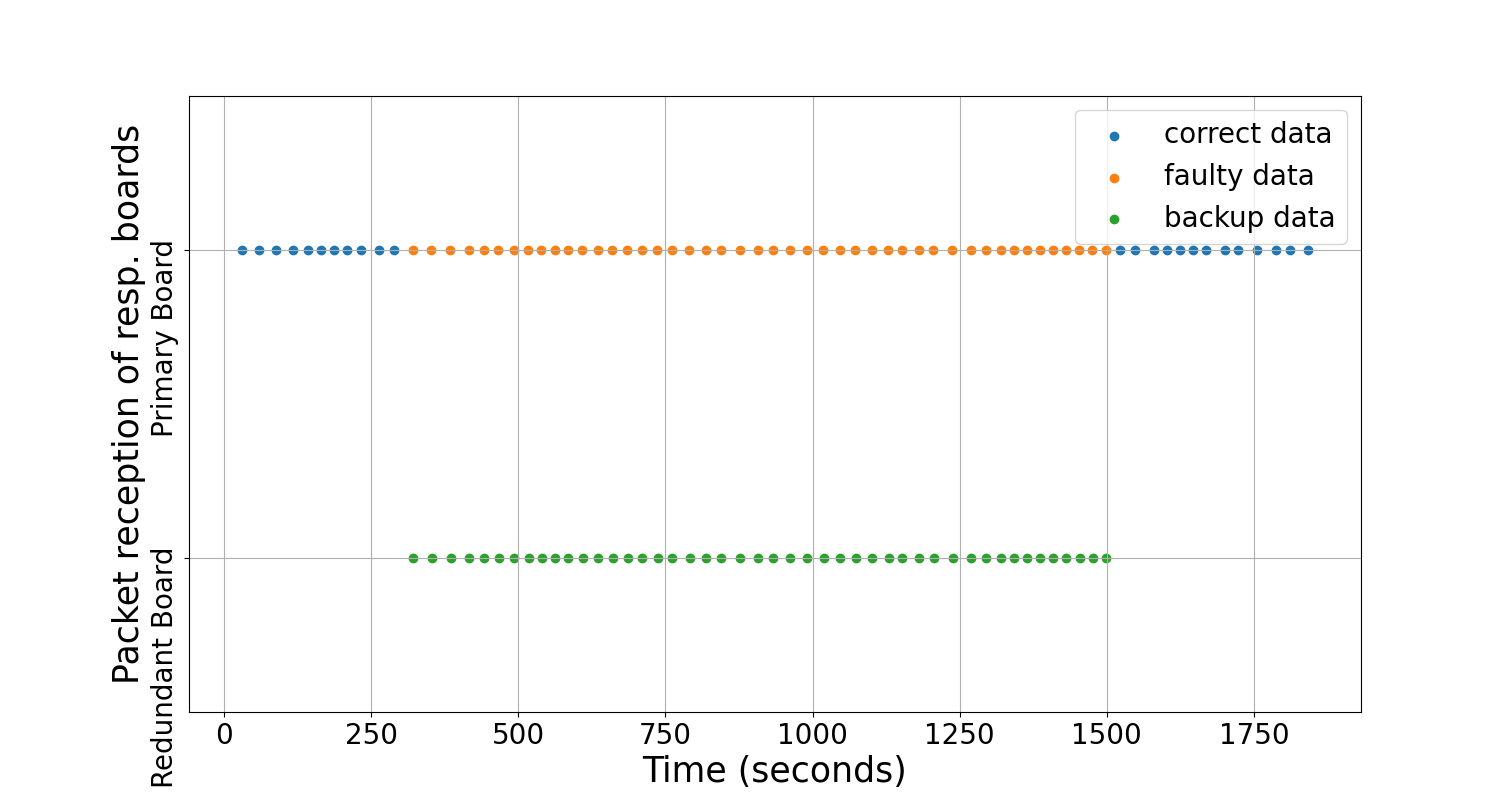}
         \caption{Control (without noise)}
         \label{fig_timing_diag_nonoise}
     \end{subfigure}
     \hfill
     \begin{subfigure}[b]{0.55\textwidth}
         \centering
         \includegraphics[width=\textwidth]{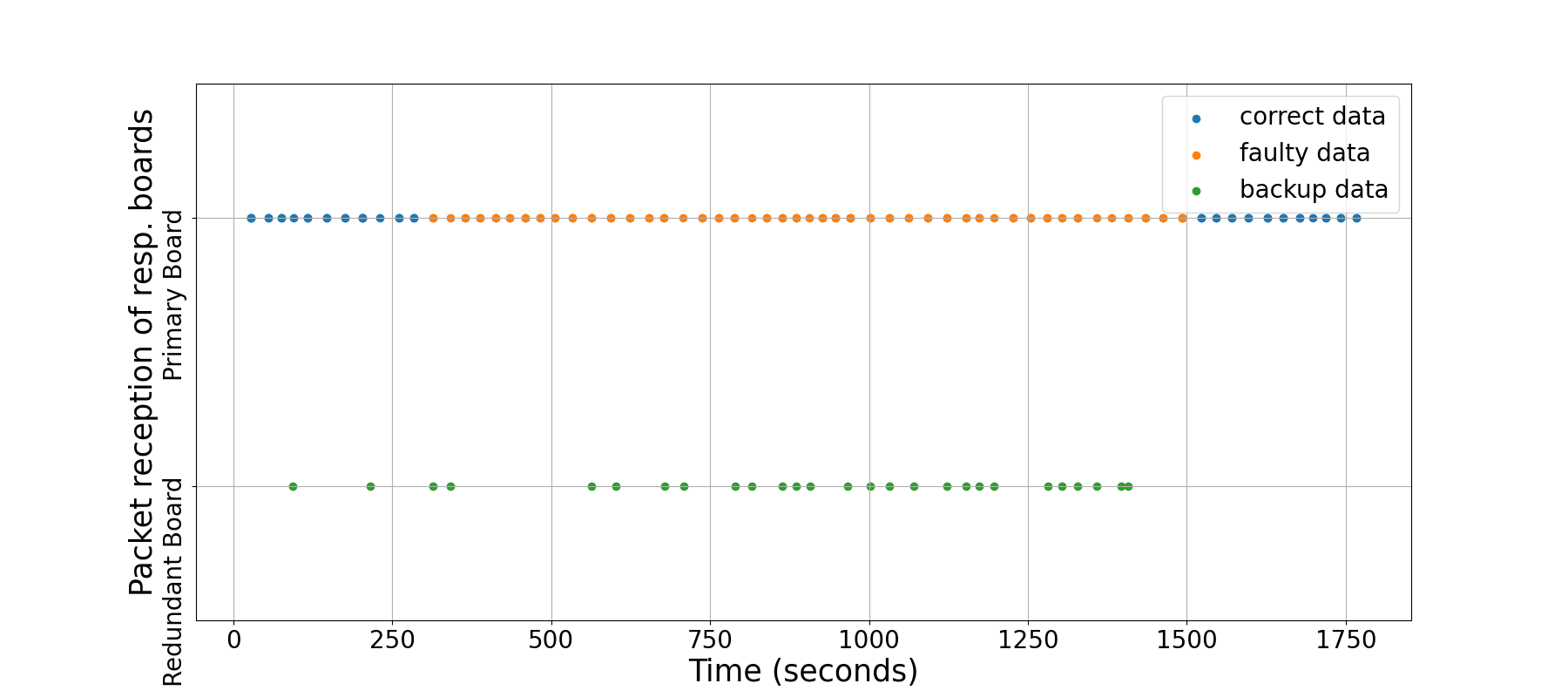}
         \caption{Actual (with noise)}
         \label{fig_timing_diag_real}
     \end{subfigure}
     \caption{Timing diagram to demonstrate the working of the system in real experiments. The figure presents all the received packets from the sensor node, and differentiates between, correct packets, faulty packets and packets from the secondary board in response to faulty packets}
\end{figure}

\begin{figure}[h!]
    \centering
    \includegraphics[width=0.5\textwidth]{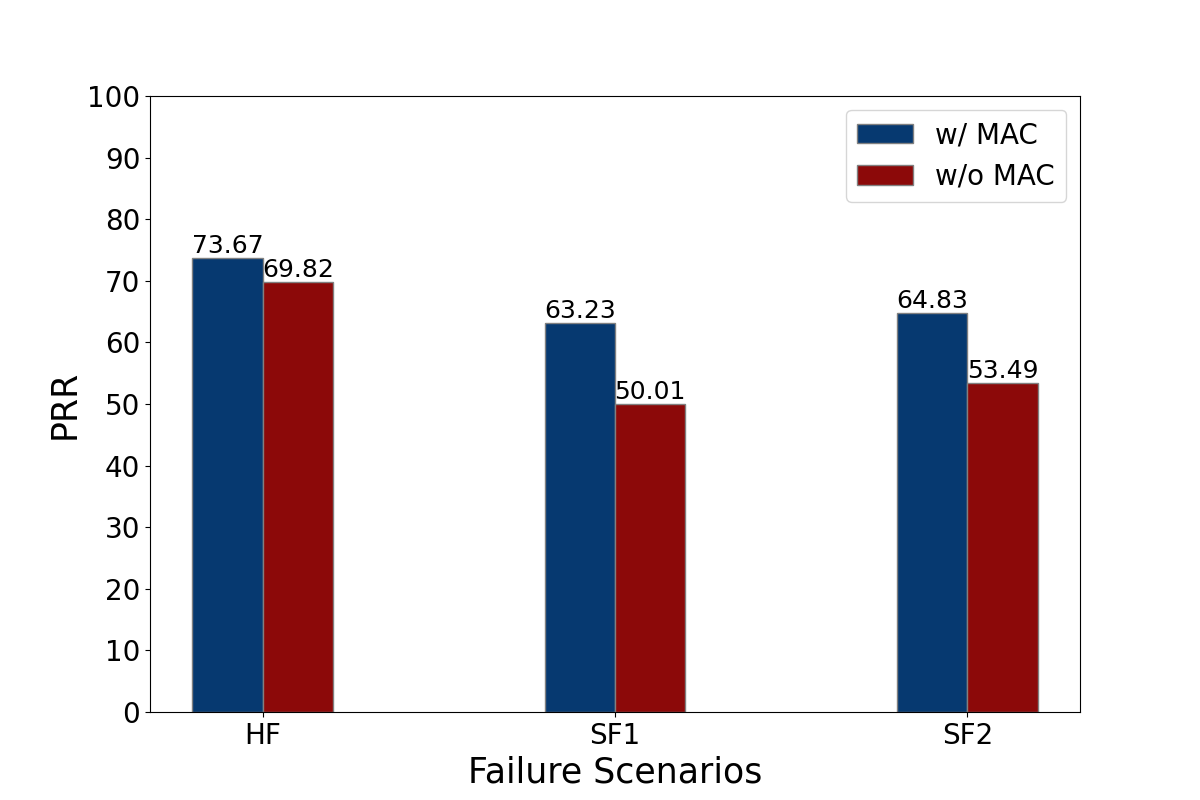}
    \caption{Influence of SARB on the packet reception of the node}
    \label{fig_mac_effect}
\end{figure}

\begin{figure}[h!]
    \centering
    \includegraphics[width=0.5\textwidth]{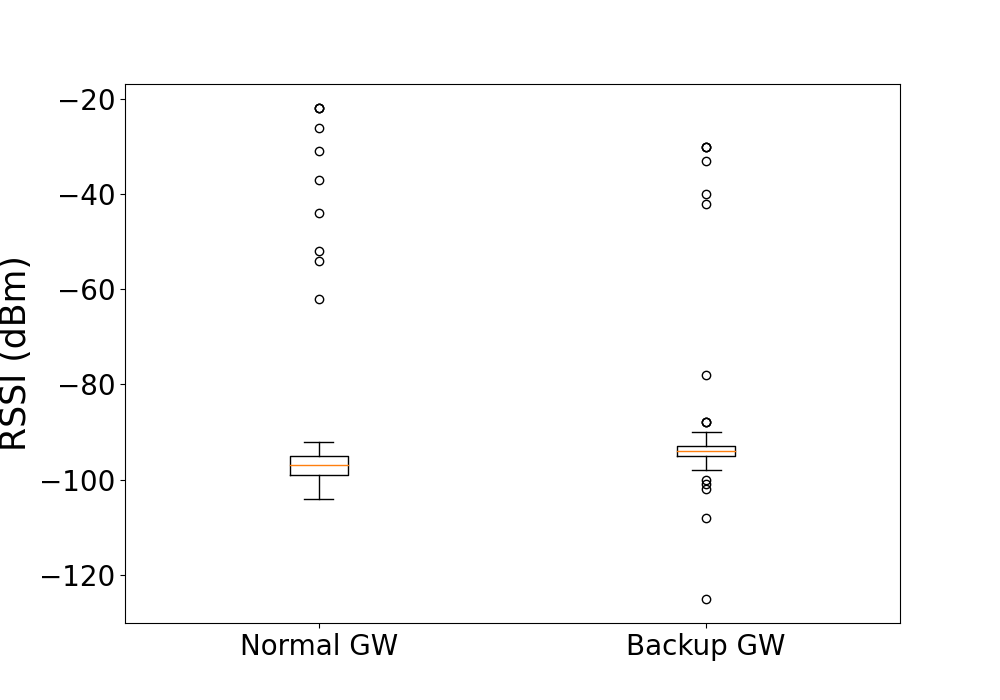}
    \caption{RSSI of the packets received by Normal Gateway Vs. Backup Gateway. }
    \label{fig_rssi_boxplot}
\end{figure}

\section{Discussion} \label{sec_discussion}

As seen in Figure \ref{fig_redundancy_prr}, the results show that by introducing redundancy in the system, we can recover packets that would otherwise be lost. In case of hard failure, no packets would be sent by the primary board. Thus the secondary board acts as a substitute. Whereas for the other two scenarios, the secondary board detects if data is faulty or incomplete and sends correct data packets. The performance is improved by $34.06\%$ for HF and by $30.42\%$ and $34.17\%$ for SF1 and SF2, respectively. For all the failure scenarios in Figure \ref{fig_redundancy_prr}, the performance of nodes without redundancy (red) is consistently compared to those with redundancy (blue). This is due to the lack of MAC implementation on the secondary nodes. The primary board can recover the packet by re-transmission, but the secondary board cannot. The efficiency of the secondary board for different failure scenarios is shown in Figure \ref{fig_detection_rate}. As the traffic increases in the sensor failure experiments, the probability of packet collision increases, leading to the loss of packets from secondary boards. 

The experiments show the effects of a noisy environment on the efficiency of the system. The effect is visualized in Figures \ref{fig_timing_diag_nonoise} and \ref{fig_timing_diag_real}. The orange points on the primary board plane represent the faulty data, and the green points on the secondary board plane represent the corresponding correct data from the secondary board. In Figure \ref{fig_timing_diag_nonoise}, the system is running in a controlled environment with no noise on the channel. All the faulty packets (orange) are detected and transmitted by the secondary board (green). On the other hand, we see some packet loss in Figure \ref{fig_timing_diag_real}, where we introduce noise on the channel. As the secondary board does not implement re-transmission, the packets that are lost cannot be recovered. One of the interesting things in Figure \ref{fig_timing_diag_real} is the two green points before $250\ seconds$. It shows us that secondary boards also help recover the packets lost due to channel congestion. Even though these packets would be re-transmitted by the primary board, they would still not meet the maximum monitoring delay constraint ($40\ seconds$). 

The MAC algorithm's influence on the system's performance is shown in Figure \ref{fig_mac_effect}. In spite of being a simple algorithm, in the absence of SARB, the performance of the system drops by $3.85\%$ in the HF scenario and by $13.22\%$ and $11.34\%$ in SF1 and SF2 scenarios, respectively.   

As seen in Figure \ref{fig_rssi_boxplot}, the RSSI values of packets received by the main gateway (Normal GW) are around -98dBm, with a few outliers with higher strength. The lower value of RSSI is due to the Automatic Gain Controller (AGC) present on the devices. However, the performance is comparable even when the gateway is placed outside the habitat (Backup GW). This demonstrates that we can use neighbouring gateways as backup receivers.

Based on all the results, we summarise how the systems react to different types of failures in Table \ref{tab_secnarios}. The system is able to recover the packets in case of Board Failure and Sensor Failure. However, in the case of anomalous data from sensors, it can detect if anomalous data exists but cannot identify where exactly the problem lies. In case of Gateway Failure, the system is able to pick up packets with the help of neighbouring gateways, but this solution is not very reliable as the PRR can drop with an increase in the distance between the neighbouring gateways. Also, this approach would work only for a wireless communication technology that uses spread spectrum modulation. Even though tackling channel congestion is not the focus of this implementation, we saw that it still helps to recover lost packets irrespective of the faults in the primary board. One scenario that cannot be tackled by the system at all is the Common Cause Failures. 

\begin{table}[]
\caption{Scope of the proposed architecture in various scenarios}
\centering
\begin{tabular}{|l|lll|}
\hline
\multirow{2}{*}{Scenarios}                                    & \multicolumn{3}{l|}{Reliability}                             \\ \cline{2-4} 
                                                              & \multicolumn{1}{l|}{Yes} & \multicolumn{1}{l|}{No} & Partial \\ \hline
Board Failure                                                 & \multicolumn{1}{l|}{\checkmark}   & \multicolumn{1}{l|}{}   &         \\ \hline
Sensor Failure                                                & \multicolumn{1}{l|}{\checkmark}   & \multicolumn{1}{l|}{}   &         \\ \hline
Sensor Anomaly                                                & \multicolumn{1}{l|}{}    & \multicolumn{1}{l|}{}   & \checkmark       \\ \hline
Gateway Failure                                               & \multicolumn{1}{l|}{}    & \multicolumn{1}{l|}{}   & \checkmark       \\ \hline
CCF                                                           & \multicolumn{1}{l|}{}    & \multicolumn{1}{l|}{\checkmark}  &         \\ \hline
\begin{tabular}[c]{@{}l@{}}Channel \\ Congestion\end{tabular} & \multicolumn{1}{l|}{}    & \multicolumn{1}{l|}{}   & \checkmark       \\ \hline
\end{tabular}
\label{tab_secnarios}   
\end{table}

\section{Applications and Future Scope} \label{sec_futurescope}
The proposed solution can be applied in different IoT applications where WiFi or cellular communication is not possible or affordable, for example, in extraterrestrial habitats, underground sensing, and monitoring of mining sites. Our proposed solution can be leveraged to increase system robustness and reliability. This system can also be used for critical applications where reliability is of utmost priority. 

One of the drawbacks of the system is that it cannot diagnose the fault that has occurred. With the proposed methods, the monitoring system is only able to mitigate the effects of various failures but does not provide any functionality to obtain debugging data or manage the nodes remotely. We will address this issue in our upcoming work. Another drawback is the use of backup boards which increases the hardware on board. However, using additional hardware cannot be avoided in life-critical applications like extraterrestrial habitats.  

Moreover, the MAC algorithm used in this work is quite simple. To make the communication of the monitoring system more reliable, a more sophisticated algorithm could be used that suits the system's requirements.

\section{Conclusion} \label{sec_conclusion}

We implemented the monitoring system for extraterrestrial habitats with secondary devices to tackle various failures in the sensor network after deployment. We evaluated the effect of secondary devices theoretically using the Continous-Time Markov Chain and experimentally using the hardware prototype of the system for the given use-case of MaMBA habitat. Based on the Markov Chain model, we can conclude that using 1 secondary device is a good trade-off between reliability and additional cost. We performed the experiment in the presence of noise on the communication channel to simulate real-life deployment and tested the system for various failure scenarios. The empirical evaluation demonstrated that the system's packet reception improves by $34.06\%$ for hard failure and by $30.42-34.17\%$ for sensor failures. We demonstrated the concept of responsive redundancy, which ensures that packets reach the destination within the defined monitoring delay in case of node failure. The secondary board not only acts as a backup for node failure but also helps to meet the monitoring requirement in case of packet loss or corruption. We also observe that using an appropriate MAC algorithm according to the application is very important for reliable communication. Selecting the most sophisticated algorithm is not always the best option as it might add unwanted overhead in communication. One of the options would be to combine existing algorithms to find the correct trade-off as per the requirements, like the SARB algorithm used in this paper. The algorithm is quite simple, but it improves the system's performance by $3.85-13.22\%$ depending on the scenario. Even though the work is motivated by extreme conditions of extraterrestrial habitats, it could also be used in other applications on Earth, like monitoring of mining sites, submarines, etc., where failure could be life-threatening. 

\section*{Acknowledgement} The authors would like to thank Ronald Mairose, ZARM, University of Bremen, for supporting us in building the hardware prototype of the system. We also extend our gratitude to our colleagues and anonymous reviewers for their feedback and suggestions.

\bibliographystyle{plain}
\bibliography{references}  






\end{document}